# The possibility of a simple derivation of the Schwarzschild metric

**Jan Czerniawski**
Institute of Philosophy, Jagiellonian University, Grodzka 52, 31-044 Krakow, Poland

E-mail: uzczerni@cyf-kr.edu.pl

**Abstract.** In spite of alleged impossibility proofs, „simple derivations" of the Schwarzschild metric, based solely on Einstein's equivalence principle and Newton's free fall velocity formula, are presented.

## 1. Introduction

Some authors claim that no simple derivation of the Schwarzschild metric is possible without the explicit use of general relativity. In their opinion, "It is the spatial-distortion aspect of gravity that ensures that too simple a derivation of the Schwarzschild metric must fail"[1]. Although they do not state it explicitly, they seem to think that this aspect is so counterintuitive that it cannot be derived in any intuitive way. However, it has long been well known that the curvature of space can be visualized very easily in terms of an appropriate distortion of measuring rods [2]. Thus, additional reasons seem to be needed.

Two other arguments for the above opinion are not so trivially inconclusive. First, the interpretation of the equivalence principle that is used in such derivations is claimed to lead to an incorrect formula for the space-time metric in the static parallel gravitational field [3]. Secondly, it is maintained that both the Schwarzschild field and its counterpart in Nordström's theory satisfy the equivalence principle; they are, however, essentially different, since only in the first of them is the empirically confirmed global bending of light present [4]. They will be examined subsequently.

## 2. Rindler's counterexample

The first alleged counterexample against the possibility of a "simple derivation" of the Schwarzschild metric, raised by W. Rindler [3], rests on a derivation of the metric in a static parallel gravitational field, which has little in common with typical "simple derivations" [5,7]. Let us therefore try to derive it more in their spirit. According to Einstein's equivalence principle [8], the influence of gravitation on phenomena in a local reference frame that is at rest in the field is equivalent to the influence of the accelerated motion of a local reference frame in which phenomena are described in the absence of gravitation. It is easy to see that the latter reduce to the length contraction, time dilation and simultaneity distortion corresponding to the velocity of the frame acquired as a result of the acceleration. Thus, if the acceleration is parallel to the axis of the coordinate $x$, the metric in such a frame acquires the form:

$$ds^2 = c^2 dt'^2 - dx'^2 - dy^2 - dz^2 \; , \tag{1}$$

where the local coordinates $x'$, $t'$ in the accelerated frame are related to the corresponding coordinates in the stationary frame by the usual differential form of the Lorentz transformation for the value of the velocity that has been reached.

Now, let us consider a local frame at rest in a static gravitational field parallel to the axis of the coordinate $x$, absent for positive values of this coordinate. The equivalence principle in the above interpretation implies that the metric in such a frame will be expressed by (1), where $x'$, $t'$ are local coordinates defined in this frame by means of physical measuring rods and clocks. Let us extrapolate the coordinate $t$ in a frame that is at rest outside the field on the area of the field in such a way that its scale remains constant through space[1] and it agrees with the local time coordinates in the frames at rest in the field with respect to simultaneity. For such a choice of this coordinate, the results of the influence of gravitation on physical objects reduce to effects analogous to length contraction and time dilation. Thus, they are given by the formulae:

$$dt' = dt\sqrt{1 - \frac{v^2}{c^2}}, \qquad (2)$$

$$dx' = \frac{dx}{\sqrt{1 - \frac{v^2}{c^2}}}. \qquad (3)$$

By substituting the above expressions into (1), we obtain:

$$ds^2 = c^2(1 - \frac{v^2}{c^2})dt^2 - \frac{dx^2}{1 - \frac{v^2}{c^2}} - dy^2 - dz^2. \qquad (4)$$

Let the quantity $v$ be given by the velocity formula for free fall in a field with gravitational acceleration $g$ antiparallel to the axis of the coordinate $x$, with the initial value zero for $x = 0$:

$$v^2 = -2gx. \qquad (5)$$

By substituting this expression into (4), we obtain [10]:

$$ds^2 = c^2(1 + \frac{2gx}{c^2})dt^2 - \frac{dx^2}{1 + \frac{2gx}{c^2}} - dy^2 - dz^2. \qquad (6)$$

Now, let us introduce new coordinates $T, X, Y, Z$ such that $X$ is related to $x$ by the formula:

$$x = \frac{g}{2c^2}X^2 - \frac{c^2}{2g} \qquad (7)$$

and $T, Y, Z$ are equal to $t, y, z,$ respectively. We then get [10]:

$$ds^2 = \frac{g^2}{c^2}X^2 dT^2 - dX^2 - dY^2 - dZ^2. \qquad (8)$$

It is easy to see that, up to a special choice of units, this is exactly what we ought to have [3], contrary to Rindler's opinion.

**3. Sexl's counterexample**
Let us consider the second of the alleged counterexamples, ascribed to R. Sexl [4]. Nordström's theory can be formulated as a metric theory of gravitation, which differs from the general theory of relativity with respect to the field equations. As a result of this difference, its metric is conformally flat, which means that in its space-times there is no global light bending, which is one of the most important predictions of general relativity. On the other hand, it may seem [11] that, as a metric theory, Nordström's theory should satisfy the equivalence principle. If so, then this principle should be insufficient for deriving the Schwarzschild metric rather than its counterpart in that theory.

Certainly, if Einstein's equivalence principle is understood as implying only that the theory of gravitation be a metric theory, such a conclusion is unavoidable. However, is this all Einstein originally meant [8]? It seems clear that any theory satisfying the equivalence principle must be a metric theory, but the converse is not as obvious. If gravitation has to be locally equivalent to (accelerated) motion of the reference frame, it seems that it should (i) affect the temporal dimension and the spatial dimension of physical phenomena parallel to the direction of the field, but (ii) leave the two other spatial dimensions unaffected. This is expressed by the approximate form (1) of the metric in a reference frame that is at rest in a gravitational field.

---
[1] This is always possible in a stationary field - see, e.g., [9], p. 120.

On the other hand, the space-time of Nordström's theory has a metric of the form [4]:

$$ds^2 = e^{2\Phi/c^2}(c^2dt^2 - dx^2 - dy^2 - dz^2) ,\qquad(9)$$

where $\Phi$ is position-dependent. It is easy to see that this implies that gravitation affects all spatiotemporal dimensions of phenomena in the same way. Thus, the forms (1) and (9) coincide, up to a constant factor, only in the trivial case when gravitation is absent. Consequently, the metric field (9) does not satisfy the equivalence principle, as Einstein seems to have understood it.

**4. The Lenz-Schiff derivation**

Let us consider a local reference frame at rest in a static spherically symmetric gravitational field. Let $r$ be the "radial" coordinate defined in the standard way and $t$ the standard time coordinate of an "infinitely" distant observer (see, e.g., [9], pp. 120 and 136). By an argument similar to that of Sec. 2, in consequence of the equivalence principle, the metric in the above-mentioned frame will have the form:

$$ds^2 = c^2 dt'^2 - dx'^2 - r^2(d\vartheta^2 + \sin^2\vartheta \, d\varphi^2) ,\qquad(10)$$

where the local spatial coordinate $x'$ in the radial direction and the local time coordinate $t'$ are related to $r, t$ by the formulae [5,7]:

$$dx' = \frac{dr}{\sqrt{1 - \frac{v^2}{c^2}}} ,\qquad(11)$$

$$dt' = dt\sqrt{1 - \frac{v^2}{c^2}} ,\qquad(12)$$

where $v$ is the velocity that a radially accelerated frame in the absence of gravitational field must have in order to be equivalent to the stationary frame in question for a given value of the coordinate $r$ (it is easy to see that the formula (12) is identical with (2)).

After the appropriate substitutions, we get a metric of the form:

$$ds^2 = c^2(1 - \frac{v^2}{c^2})dt^2 - \frac{dr^2}{1 - \frac{v^2}{c^2}} - r^2(d\vartheta^2 + \sin^2\vartheta \, d\varphi^2) .\qquad(13)$$

Let us assume that $v$ is given by the Newtonian formula for free fall with the initial velocity zero "at infinity", i.e., it is equal to the escape velocity in Newton's gravitation theory:

$$v^2 = \frac{2kM}{r} ,\qquad(14)$$

where $k$ is the gravitational constant and $M$ is the mass of the source of the gravitational field. Finally, by substituting (14) to (13), we get the form of the Schwarzschild metric:

$$ds^2 = c^2(1 - \frac{2kM}{rc^2})dt^2 - \frac{dr^2}{1 - \frac{2kM}{rc^2}} - r^2(d\vartheta^2 + \sin^2\vartheta \, d\varphi^2) .\qquad(15)$$

**5. The river model derivation**

What might make one a bit uneasy about the Lenz-Schiff derivation is the unclear status of $v$. It cannot, as one might think, be interpreted as the velocity of a frame falling in a gravitational field, understood as the first derivative of the coordinate $r$ with respect to the time coordinate $t$. As we have observed in the previous section, it is rather the velocity of a frame in the absence of gravitation.

Thus, its meaning is specified in circumstances different from those in which the metric is derived. One might therefore regard the success of the derivation as a mere coincidence.

Fortunately, there is another derivation, in which the meaning of $v$ is clear. Let us assume that gravitation reduces to some motion, in a Euclidean space, of a "substratum", in which special relativity holds locally. This means that in a resting frame everything goes on as though it moved and the "substratum" rested. Thus, in such a river model [12] of gravitation, Einstein's equivalence principle is trivially satisfied.

Let in a spherically symmetric field the "substratum" fall radially, obeying Newton's law of free fall, and let $v$ be the velocity of its falling. Finally, let in comoving frames the spatiotemporal dimensions of physical phenomena remain the same as if the "substratum" rested. This means, in particular, that a comoving clock synchronized with a clock resting "at infinity" will retain synchrony with that clock during its falling. Thus, a time coordinate $T$ can be defined by such clocks synchronized in this way. Moreover, the distances between simultaneous events measured by comoving rods are just the distances in Euclidean space. It is easy to see that, in consequence of the above assumptions, the space-time metric in a comoving frame is locally approximated by (10), in which $r$ is interpreted as the usual radial coordinate and the local spatial coordinate in the radial direction $x'$ and the time coordinate $t'$ are related to $r, T$ by the approximate formulae of the differential form of the Galilean transformations:

$$dr' = dr - vdT , \qquad (16)$$
$$dt' = dT , \qquad (17)$$

where $v$, interpreted as the velocity of the origin of the frame, is a function of its time-dependent coordinate $r_0$. By substituting (16) and (17) into (10), we get:

$$ds^2 = c^2 dT^2 - (dr - vdT)^2 - r^2 (d\vartheta^2 + \sin^2 \vartheta d\varphi^2) , \qquad (18)$$

or, equivalently:

$$ds^2 = c^2(1 - \frac{v^2}{c^2})dT^2 + 2vdTdr - dr^2 - r^2(d\vartheta^2 + \sin^2 \vartheta d\varphi^2) . \qquad (19)$$

In this way, the local coordinates $x'$, $t'$ have been eliminated. The approximate expression (19) is the more accurate the less $r$ differs from $r_0$. Thus, if we substitute $r$ for $r_0$ in $v$, this expression for the metric becomes exact. Now, let us assume that the velocity $v$ satisfies the Newtonian free fall law, expressed by (14). We obtain the following form of the metric:

$$ds^2 = c^2(1 - \frac{2kM}{rc^2})dT^2 - 2\sqrt{\frac{2kM}{r}}dTdr - dr^2 - r^2(d\vartheta^2 + \sin^2 \vartheta d\varphi^2) , \qquad (20)$$

found for the first time by A. Gullstrand [13] and known as the Painlevé-Gullstrand metric [14], or the Painlevé-Gullstrand-Lemaitre metric (cf.. [15] and related references therein), although it seems that in the works of Painlevé and Lemaitre this precise form does not appear.

In the river model the metric (20) was derived for the first time by A. Trautman [16]. It might seem very different from the Schwarzschild metric. However, it is well-known to be just the Schwarzschild metric in non-standard coordinates. It can be turned into the usual form by the following transformation of the time coordinate [14,15]:

$$dT = dt + \frac{\sqrt{\frac{2kM}{r}}}{c^2(1 - \frac{2kM}{rc^2})} dr . \qquad (21)$$

It is worth noting that the metric in a constant field given by the expression (6) was originally derived [10] in the Painlevé-Gullstrand form:

$$ds^2 = c^2(1 + \frac{2gx}{c^2})dT^2 - 2\sqrt{-2gx}dTdx - dx^2 - dy^2 - dz^2 \qquad (22)$$

and then transformed into (6) by a transformation given by the formula:

$$dT = dt + \frac{\sqrt{-2gx}}{c^2(1 + \frac{2gx}{c^2})}dx \; , \qquad (23)$$

analogous to (21).

## 6. Discussion

One might think that the formula (20) reveals the "true" form of the Schwarzschild metric. Yet such a conclusion would be premature. Let us return to (14). It has two solutions with different signs, which apply to free fall with the "initial" value zero at infinity and to free escape with the "final" value zero. The solution with a negative sign was used in the derivation of Sec. 5. Why not, then, assume that the "substratum", instead of flowing inwards, flows radially outwards, obeying the Newtonian law of free escape?

Surprisingly, this also works. Following steps completely analogous to that derivation, we now get the metric with the form:

$$ds^2 = c^2(1 - \frac{2kM}{rc^2})dT^2 + 2\sqrt{\frac{2kM}{r}}dTdr - dr^2 - r^2(d\vartheta^2 + \sin^2\vartheta d\varphi^2) \; . \qquad (24)$$

This is just the other version of the Painlevé-Gullstrand metric [14,15], discovered independently by A. Gullstrand [13] and P. Painlevé [17], and rediscovered by G. Lemaitre [18].

There is yet another river model derivation, found by W. Laschkarew [20]. Instead of the time coordinate $T$, one can introduce the coordinate $t$ in the standard way applicable in static fields (cf. Sec. 2 above). Then all the steps of the derivation of Sec. 4 follow, with the only difference that this time the expressions (11) and (12) have a very simple interpretation, as expressing the length contraction and time dilation, respectively, which result from the stationary frame's motion relative to the "substratum".

The river model may seem attractive as it is very intuitive. Nevertheless, it has some weak points. One of them is the above-mentioned duality of the river model derivations of the Schwarzschild metric. Next, one can observe that, if the assumptions of the river model are adopted, then the effects expressed by (11) and (12) are to be expected, but the reverse is not true. Thus, the assumption that there is some real "substratum" flow seems unnecessary for deriving the general relativistic metric.

What is more, an excessively realistic approach to the river model encounters serious difficulties. First, if the homogeneity of the "substratum" is assumed, then its flow does not seem to satisfy the continuity equation [14]. Otherwise, one would have to explain why its possible inhomogeneities do not affect physical rods and clocks. Next, although the model has proved successful in deriving some interesting special cases of general relativistic metrics [10,16], the most surprising being the Kerr metric [12], its applicability to all physically reasonable solutions of the Einstein equations remains to be shown. Thus, some reservations with respect to it seem in order.

**References**

[1] Gruber R P, Price R H, Matthews S M, Cordwell W R and Wagner L F 1988 The impossibility of a simple derivation of the Schwarzschild metric *Am. J. Phys.* **56** 265-269
[2] Einstein A 1921 *Geometrie und Erfahrung* (Berlin: Springer)
[3] Rindler W 1968 Counterexample to the Lenz-Schiff argument *Am. J. Phys.* **36** 540-544
[4] Ehlers J and Rindler W 1997 Local and global light bending in Einstein's and other gravitational theories *Gen. Relativ. Gravit.* **29** 519-529
[5] Lenz W 1944 unpublished work cited in [6] p. 313
[6] Sommerfeld A 1967 *Electrodynamics* (*Lectures on Theoretical Physics* vol. 3) (New York: Academic Press)
[7] Schiff L I 1960 On experimental tests of the general theory of relativity *Am. J. Phys.* **28** 340-343
[8] Einstein A 1911 Über den Einfluss der Schwerkraft auf die Ausbreitung des Lichtes *Ann. Physik* **35** 898-908
[9] Rindler W 1977 *Essential Relativity* 2nd ed. (New York: Springer)
[10] Visser M 2003 Heuristic approach to the Schwarzschild geometry *Preprint* gr-qc/0309072
[11] Will C M 1987 Experimental gravitation from Newton to Einstein *300 Years of Gravitation* ed. S W Hawking and W Israel (Cambridge: Univ. Press) pp. 80-127
[12] Hamilton A J S and Lisle J P 2004 The river model of black holes *Preprint* gr-qc/0411060, to appear in *Am. J. Phys.*
[13] Gullstrand A 1922 Allgemeine Lösung des statischen Einkörpersproblem in der Einsteinschen Gravitationstheorie *Arkiv Mat. Astron. Fys.* **16** (8) 1-15
[14] Visser M 1998 Acoustic black holes: horizons, ergospheres and Hawking radiation *Class. Quantum Grav.* **15**, 1767-1791
[15] Schützhold R 2001 On the Hawking effect *Phys. Rev.* D **64** 024029
[16] Trautman A 1966 Comparison of Newtonian and relativistic theories of space-time *Perspectives in Geometry and Relativity* ed. B Hoffman (Bloomington: Indiana Univ. Press) pp. 413-425
[17] Painlevé P 1921 La Mécanique classique et la théorie de la relativité *C. R. Acad. Sci. (Paris)* **173** 677-680
[18] Lemaitre G 1933 L'univers en expansion *Ann. Soc. Sci. (Bruxelles)* A **53**, 51-85
[19] Lemaitre 1997 The expanding universe *Gen. Relativ. Gravit.* **29** 641-680, Eng. transl. of [18]
[20] Laschkarew W 1926 Zur Theorie der Gravitation *Z.Physik* **35** 473-476